
\magnification=\magstep1
\baselineskip=24 true pt
\hsize=33 pc
\vsize=40 pc
\bigskip
\bigskip
\rightline {NI94023}
\rightline {October, 1994}
\centerline {\bf FOUR DIMENSIONAL STRING-STRING SOLUTIONS AND}
\centerline {\bf SYMMETRIES OF STRING EFFECTIVE ACTION}
\vfil
\centerline {{\bf Jnanadeva
Maharana}\footnote\dag{Permanent address:Institute of Physics, Bhubaneswar
751005, India}}

\bigskip
\centerline {\it { Isaac Newton Institute for Mathematical Sciences} }
\centerline {\it { Cambridge, CB3 0EH, U.K.}}
\vfil
\centerline {\bf Abstract}

A string  action is considered in four spacetime dimensions
which is obtained  by dimensionally reducing the ten dimensional
effective action. The
equations of motion admit string like solutions. The symmetry properties
of the four dimensional action is discussed. It  is shown that  new background
configurations can be generated by implementing suitable $O(d,d)$
transformations.

\vfil
\eject


Recently, considerable attention has been focussed in investigating the
target space symmetries of string theory; notable among them being duality
and noncompact global symmetries. The target space duality[1] is one of the
important symmetries of string theory. If we consider a
compactified string on a
circle of radius $R$, the spectrum of the string remains invariant under the
transformation $ R \rightarrow {{\lambda}^2 \over R}$ ( with ${\lambda}^2} =
2 \alpha' \hbar$). The consequences of duality have been explored in various
directions. The global $O(d,d)$ symmetry of the string effective action, for
the cosmological case, was discovered by Meissner and
Veneziano[2] and it was shown
subsequently, how one can generate nontrivial cosmological geometries by
applying these transformations[3,4].
Sen and
his collaborators[5] extended these  results to the case when
the backgrounds are
independent of $d$-coordinates and generated new backgrounds by applying the
$O(d,d)$ transformations. The dimensional reduction technique
due to  Scherk and
Schwarz[6] was adopted by Schwarz and Maharana[7] to demonstrate the $O(d,d)$
invariance of the effective action when $d$-coordinates are toroidally
compactified and the background fields are independent of these coordinates.

There has been some progress, in the recent past, to understand
the nonperturbarive aspects of string theory. A clue is provided
from the string/fivebrane duality conjecture[8,9]. In the
critical, $D=10$, dimension strings ( one dimensionalobjects )
are dual to fivebranes, which are extended objects with five spatial
dimensions. A considerable attention has been focussed on the
study of $D=10$ heterotic strings and the $D=10$
fivebranes[9-13] and the latter is the dual counter part
of the former. Moreover, the strong/weak coupling duality was
proposed[9] in analogy with with the strong/weak coupling
conjectured in super Yang-Mills theories[14,15]. The strong/weak
coupling duality can be viwed in the frame work of Poincare duality,
 where it was shown that if the string coupling constant is
given by $exp({\phi_0})$, the corresponding fivebrane coupling
constant is $exp(-{{\phi_{0}} \over {3}})$, where $\phi_0$ is
the VEV of dilaton. Thus, weak coupling in string theory
corresponds to strong coupling in fivebrane and vice versa. It was pointed out
by Strominger[9] that the fivebrane appears as, a zerobrane, a
onebrane, or a twobrane after toroidal compactification to four dimensions. The
point like and extended objects manifest themselves after compactification
depending on how the
fivebrane wraps around the compactified dimensions. Therefore, it seems quite
natural to seek these solitonic objects as solutions of the four dimensional
string effective action. The zerobrane, onebrane and twobrane correspond to
monopole, string and domain wall solution as was demonstrated in by Duff and
his collaborators [17]. There exist fundamental string solutions[18,19] with
many interesting features. The solitonic string solution admits a duality
transformation
where the 'dilaton' and 'axion' coming from the moduli can be grouped together
and are paramtrised by $SL(2,Z)$ and the corresponding duality is known as the
T-duality. It is worthwhile to point out that similar duality transformations
were discovered by Khastgir and Maharana[20] when they were seeking Eucleadian
wormhole solutions of four dimensional effective action. In the  case of
the fundamental string, the dilaton and the axion are similarly  gouped into a
multiplet and the $SL(2,Z)$ transformation  relates the strong and weak
coupling regimes of the theory and therefore, it is hoped that one will be able
to compute  effects beyond the perturbation theory[21,22]. Therefore, under
string/fivebrane duality, the $SL(2,Z)$ strong/weak coupling duality gets
exchanged with $SL(2,Z)$
target space duality as observed by Schwarz and Sen[22] and Binetruy[23]. The
prospect of exploring nonperturbative aspects of string theory has stimulated
investigations in the recent past[23-25].

The purpose of this note is to explore further the solitonic strings which
arise as solutions of the string effective action and generate new background
configurations by applying the noncompact symmetry transformations. We shall
show that several interesting configurations emerge in this process. First we
shall recapitulate the essential results of the dimensional reduction technique
adopted by Schwarz and the author to show the invariance properties of the
reduced string effective action. Next, we recall the results of Duff and
collaborators and demonstrate that the action considered by them is invariant
under noncompact global symmetry transformations and target space duality
transformation. Subsequently, we shall generate new backgrounds which satisfy
equations of motion.

We now recapitulate the essential steps to construct the string effective
action. Let us consider  the evolution of the string in the background of
its massless  excitations.
The conformal
invariance of the worldsheet action demands the vanishing of the
$\beta$-functions
associated with the backgrounds. In other words,these conditions are the
equations of motion to be satisfied by the backgrounds to ensure conformal
invariance of the theory.
The tree level string effective action, involving
only the massless excitations, can be constructed in such a way that
the equations of motion, derived from this effective action exactly
coincide with the vanishing of the $\beta$- functions.

 Let us begin by recalling  the main results
of ref.7   and set the notations. The
bosonic part of the effective action in
 $\hat D=D+d$ Euclidean dimensions ($\hat D=26,10$ for bosonic,
fermionic critical cases respectively) is,

$$\hat S = \int d^{\hat D}x~ \sqrt{ \hat g}~ e^{-\hat\phi}
\big [\hat R
(\hat g) + \hat g^{\hat \mu \hat \nu} \partial_{\hat \mu} \hat\phi
\partial_{\hat \nu} \hat\phi - {1 \over 12} ~ \hat H_{\hat \mu \hat \nu
\hat \rho} ~
\hat H^{\hat \mu \hat \nu \hat \rho}\big ].\eqno (1)$$

\noindent $\hat H$ is the field strength of antisymmetric tensor and $\hat
\phi$ is the dilaton. Here we have set all the nonabelian gauge field
backgrounds to zero. When the backgrounds are independent of the `internal'
coordinates $y^{\alpha}, \alpha=1,2..d$ and the internal space is
taken to be torus, the metric $\hat g_{\hat \mu \hat \nu} $
 can be decomposed as

$$\hat g_{\hat \mu \hat \nu} = \left (\matrix {g_{\mu \nu} +
A^{(1)\gamma}_{\mu} A^{(1)}_{\nu \gamma} &  A^{(1)}_{\mu \beta}\cr
A^{(1)}_{\nu \alpha} & G_{\alpha \beta}\cr}\right ),\eqno (2)$$

\noindent where $G_{\alpha \beta}$ is the internal metric and $g_{\mu\nu}$,
the $D$-dimensional space-time metric, depend on the coordinates $x^{\mu}$.
The dimensionally reduced action is,

$$\eqalign S =& \int d^Dx \sqrt {g}~ e^{-\phi}
\bigg\{ R + g^{\mu \nu}
\partial_{\mu} \phi \partial_{\nu} \phi -{1\over 12}H_{\mu \nu \rho} ~
H^{\mu \nu \rho}\cr
&+ {1 \over 8} {\rm tr} (\partial_\mu M^{-1} \partial^\mu
M)- {1 \over 4}
{\cal F}^i_{\mu \nu} (M^{-1})_{ij} {\cal F}^{\mu \nu j} \bigg\}}.
\eqno (3)$$

\noindent Here $\phi=\hat\phi-{1\over 2}\log\det G$ is the shifted dilaton.
$$H_{\mu \nu \rho} = \partial_\mu B_{\nu \rho} - {1 \over 2} {\cal A}^i_\mu
\eta_{ij} {\cal F}^j_{\nu \rho} + ({\rm cyc.~ perms.}),\eqno (4)$$
\noindent ${\cal F}^i_{\mu \nu}$ is the $2d$-component vector of field
strengths
$${\cal F}^i_{\mu \nu} = \pmatrix {F^{(1) \alpha}_{\mu \nu}
\cr F^{(2)}_{\mu \nu \alpha}\cr} = \partial_\mu {\cal A}^i_\nu - \partial_\nu
{\cal A}^i_\mu \,\, ,\eqno (5)$$
\noindent $A^{(2)}_{\mu \alpha} = \hat B_{\mu \alpha} + B_{\alpha \beta}
A^{(1) \beta}_{\mu}$ (recall $B_{\alpha \beta}=\hat B_{\alpha \beta}$), and
the $2d\times 2d$ matrices are
$$M = \pmatrix {G^{-1} & -G^{-1} B\cr
BG^{-1} & G - BG^{-1} B\cr},\qquad \eta =  \pmatrix {0 & 1\cr 1 & 0\cr}
\, .\eqno (6)$$

\noindent The action (3) is invariant under a global $O(d,d)$ transformation,

$$M \rightarrow \Omega^T M \Omega, \qquad \Omega \eta \Omega^T = \eta, \qquad
{\cal A}_{\mu}^i \rightarrow \Omega^i{}_j {\cal A}^j_\mu, \qquad {\rm
where} \qquad
\Omega \in O(d,d). \eqno (7)$$
\noindent and the shifted dilaton, $\phi$, remains invariant
under the $O(d,d)$ transformations. We mention in passing
that $ M \rightarrow M^{-1} $ under the duality transformation.
Note that $M\in O(d,d)$ also and $M^T\eta M=\eta$. The
background equations of motion can be derived from (11). The classical
solutions of string effective action correspond to different string vacua
and are given by solutions for $M$,$\cal F$ and $\phi$. Thus, when one chooses
backgrounds satisfying equations of motion these backgrounds
correspond to vacuum
configurations of the string theory.

In what follows, we we shall briefly recall string/string solutions obtained
by Duff et al[17] from the string effective action. The massless bosonic fields
relevant for our discussion are: graviton, antisymmetric tensor and the
dilaton.
One starts from the $10$-dimensional heterotic string effective action. In
order to derive the string like solution, it is assumed that all backgrounds
are
independent of the coordinates $x^3, x^4, x^6, x^7, x^8$ and $x^9 $. Therefore,
we may consistently assume that these coordinates are compactified on a six
dimensional torus. The four dimensional space is labelled by coordinates  $
x^0, x^1, x^2$,
and $x^5$. The metric is taken to be flat along the directions $6,7,8$ and $9$
and therefore the compactification of these dimensions is trivial for the
metric. While going from six to four dimensions, it is assumed that $ g_{33} =
g_{44} = e^{-2\sigma}$. All the components of the antisymmetric tensor field
is assumed to vanish except the component $B_{34}$ and therefore, the only
surviving field strength is $H_{\rho 34} = \partial _{\rho}B_{34}$, where
$\rho = 0,1,2, 5$.

 Now we present the tree level string effective action in the form used in ref.
17  and discuss the solutions. Next, we shall identify the backgrounds
in the form suitable for the $O(d,d)$ transformation.

$$\aligneq S_4 = \int d^4x \sqrt { -g} e^{-2\Phi - 2\sigma} \bigg\{ R +
4(\partial \Phi)^2 + 8 \partial \sigma .\partial \Phi + 2(\partial \sigma)^2 -
{1 \over 2}
e^{4\sigma} (H_{\rho 34})^2 \bigg\} \eqno (8) $$

\noindent The backgrounds satisfying the equations of motion are

$$ e^{2\Phi} = e^{-2\sigma} = - e^{2 {\Phi}_0} ln(\vert \bf x - \bf a \vert)
\eqno (9) $$

$$ ds^2 = - dt^2 + dx_5^{2} + e^{2\Phi}( dx_1^{2} + dx_2^{2}) \eqno (10) $$

$$ H_{i34} = \pm \epsilon_{ij} \partial_j e^{2 \Phi}, i,j = 1,2  \eqno (11) $$

\noindent Here $\Phi$ is the dilaton field,and  $\Phi_{0}$ denotes the constant
asymptotic value of the dilaton. The solution corresponds to a single string
located at $\bf a$ and $\bf x$ is the two dimensional vector with coordinates
$x_1 $ and $x_2$. It is evident that the metric only depends on the magnitude
of the vectors $x_1$ and $x_2$ and similarly the field $B_{34}$ depends on the
magnitude which we denote as $r = ( x_1^{2} + x_2^{2})^ {1 \over 2}$ from now
on.

In order to establish the correspondence with the conventions of ref.7
, we first note that the dilaton field considered in reference 17 is
such that $2\Phi = \hat  \phi $ as is evident from comparision of the form of
effective actions; equations 3 and 7. Next, we see that the shifted dilaton
after this scaling
is $\phi = \hat \phi + 2\sigma$. Indeed, we notice that the string solution
corresponds to vanishing value of the  shifted dilaton.

It is worthwhile to point out some similarities between the form of the moduli
space chosen by Khastgir and Maharana[20] when they obtained wormhole solutions
of the string effective action and the string solutions of Duff et al. In the
case of KM, they were  seeking Eucleadian spherically symmetric wormhole (
rotations in the Euclidean space ) solutions and the internal $6 \times 6$
dimensional metric
$G$ was expressed as $3$ blocks of $2 \times 2$ matrix and each of the $2
\times 2$ matrix has the form similar to the one chosen by Duff et al.[17] to
obtain their
string solutions. Moreover, it was  shown in ref. 20  that the 'internal
dilaton', denoted
by $\sigma$ (in the notations of Duff et al. see eq.(9))  and the internal
'axion' can be grouped into a
multiplet parametrizing the coset $SL(2,R)$ and the reduced effective action
can
be expressed in a manifestly $SL(2,R)$ invariant form. Indeed, Duff et al.[17]
have also
discovered similar symmetry structure for their four dimensional
effective action as we mentioned earlier. We were able to generate new wormhole
solutions by
 implementing the $SL(2,R)$ transformations and duality transformations[20],
which transforms a
wormhole with global charge $Q_j$ to one with the charge $1 \over {Q_j}$. This
duality transformation is precisely the T-duality of ref.[17].

Therefore, it is natural to expect that one can generate new string solutions
by implementing suitable $O(d,d)$ transformations on the known solution. In
order to facilitate these transformations, we shall identify the background
fields we would like to rotate and specify the transformations. It is more
convenient to express the metric in two dimensional polar coordinates, $r$ and
$\theta$. Thus, we rewrite eq.(9) as

$$ ds^2 = -dt^2 + dx_5^{2} + e^{\phi} ( dr^2 + r^2 d{\theta}^2) \eqno (12) $$

\noindent and all the backgrounds depend only on the coordinate $r$.

Let us first consider a rotation involving the coordinate $\theta$ and $x_3$.
If we look at the components of the metric and the antisyymetric tensor in
this space we note that the only nonzero components are $g_{\theta \theta}$ and
$g_{33}$, $g_{\theta 3} = 0$ and $B_{\theta 3} = 0$ and $B_{33}
= 0$ due to the
antisymmetry property. Now, we are in a position to construct the $M$-matrix
as defined in eq.(6) and perform an $O(2,2)$ rotation, which were called
 'boosts' by Gasperini, Maharana and Veneziano[3,4] and we
shall use the same nomenclature for these trnsformations. It is
easy to see that, for the problem at hand, the $2 \times 2$
$M$-matrix is diagonal since the
 relevant $B$ field is zero.

$$ M = \pmatrix { G^{-1} & 0 \cr 0 & G\cr} \eqno (13) $$

\noindent where,

$$ G = \pmatrix { r^2 e^{2\Phi} & 0 \cr 0 & e^{2\Phi}\cr}, \eqno (14) $$

 The $O(2,2)$ matrix is parametrized the boost angle $\gamma$ and takes the
 following form

$$ \Omega (\gamma) = {1\over 2} \pmatrix {1 + c & s & c - 1 & -
s \cr - s & 1 - c & - s & 1 +c \cr c - 1 & s & 1 + c & - s \cr s
& 1 + c & s & 1 - c \cr} \eqno (15) $$

\noindent Here we have adopted the notation $c = cosh \gamma $ and $s =
 sinh \gamma$ and $\gamma$ lies between $0$ and infinity. Now,
using the transformation property

$$ M(\gamma) = \Omega^T M \Omega \eqno (16) $$

Thus, we see that the matrix $M(\gamma )$ will have nonzero off
diagonal elements.
Therefore, we shall be able to compute the form of new
background field obtained from the original ones through the
above transformations. After some simple
calculations, we arrive at the following form

$$ G' = \pmatrix { {(c-1)e^{-2\Phi} + (c+1)r^2 e^{2\Phi}} \over
{(c+1) + (c-1)r^2} & - {{ s( r^2 e^{2\Phi}+e^{-2\Phi}) \over
{(c+1) + (c-1)r^2}}} \cr - {{ s(r^2 e^{2\Phi} + e^{-2\Phi}) \over
{(c+1) + (c-1)r^2}}} & {(c+1) e^{-2\Phi} + (c-1)r^2 e^{2\Phi}}
\over {(c+1) + (c-1)r^2}} \cr} \eqno (17) $$

\noindent The first element corresponds to the $\theta - \theta
$ component of the metric and second element is $\theta - 3$
element and the metric is indeed
symmetric.  It is important to note that the form of the metric
has changed under the transformation. The antisymmetric tensor
field, which had all vanishing components inthe initial
configuration, takes the form

$$ B' = \pmatrix { 0 & {s(1 + r^2)} \over {(c+1) + (c-1)r^2} \cr
-{{s(1+r^2)} \over {(c+1) + (c-1)r^2}} & 0 \cr} \eqno (18) $$

\noindent We find that new antisymmetric tensor field has nontrivial components
, which cannot be gauged transformed to zero, and the
corresponding field strength $H_{r\theta 3}$ is nonzero since
the $B'$ depends only on the radial coordinate, $r$.

We recall that the metric components $g_{33}$ and $g_{44}$
depend only on $r$ since $\sigma$ carries only r-dependence.
Moreover, the metric as reexpressed in eqn. (12) also depends on
$r$ only. Therefore, we can generate several different types of
backgrounds through the choice of the boost matrices. We shall
present
here another transformation which involves two spacetime coordinates, $\theta$
and $x_5$ and one internal coordinate $x_3$. On this occasion the $G$-matrix is
a $3\times 3$ diagonal matrix. Thus one can generate new
backgrounds by $O(3,3)$ transformations. Let us consider a
rotation on the $\theta -x_5$ plane and
the corresponding transformation matrix is given by

$$ \tilde \Omega = {1 \over 2} \pmatrix {1+c' & 0 & s' & c'-1 &
0 & -s' \cr 0 & 2 & 0 & 0 & 0 & 0 \cr -s' & 0 & 1-c' & -s' & 0 &
1+c' \cr c'-1 & 0 & s' & 1+c' & 0 &  -s' \cr 0 & 0 & 0 & 0 & 2
&0 \cr s' & 0 & 1+c' & s' & 0 & 1-c' \cr} \eqno (19) $$

\noindent where $c' = cosh{\gamma '}$ and $s' = sinh{\gamma '}$,
and $\gamma '$
is the boost angle.
The transformed $M$-matrix is again obtained from the relation
$\tilde M(\gamma ) =
\tilde \Omega^ T M \tilde \Omega$ and then $\tilde G$ and
$\tilde B$ are obtained by a lengthy but straight forward
calculation.
We present the form of the
background field below.

$$ \tilde G = {1\over 2} \pmatrix { {\{ {{s'e^{-\Phi}} \over
{r}} + s'r^ e^{\Phi}\}^2 + 4 } \over {{\cal D}^2} & 0 & -{{s'\{{
{(c'+1)e^{-2\Phi}} \over {r^2}} + (c'-1)r^2 e^{2\Phi} + 2c' \}}
\over {{\cal D}^2}} \cr 0 & e^{2\Phi} & 0 \cr -{{s'\{
{{(c'+1)e^{-2\Phi}} \over {r^2}} + (c'-1)r^2 e^{2\Phi} + 2c' \}}
\over {{\cal D}^2}} & 0 & 1 \cr} \eqno (20) $$

\noindent whereas the antisymmetric field $\tilde B$ has the form

$$ \tilde B = \pmatrix {0 & 0 & { {s'\{{{(c'+1)e^{-2\Phi}} \over
{r^2}} + (c'-1)r^2 e^{2\Phi} + 2c' \}} \over {{\cal D}^2}}  \cr 0
& 0 & 0 \cr -{{s'\{{{(c'+1)e^{-2\Phi}} \over {r^2}} + (c'-1)r^2
e^{2\Phi} + 2c' \}} \over {{\cal D}^2}} & 0 & 0 \cr} \eqno (21)
$$

\noindent where ${\cal D} = {{(c'+1)e^{-\Phi}} \over {r}} + (c'-1)r e^{\Phi}$.
Again we observe that now the metric has developed nonzero $\theta-5$ and
$5-\theta$ components and it is symmetric. Similarly, $\tilde B$ has nontrivial
components along
$\theta-5$ and $5-\theta$ direction.

So far we have confined our attentions to discuss how the backgrounds $G$ and
$B$ transform under $O(d,d)$ transformations to give new backgrounds. Let us
consider how the dilaton is transformed under these rotations. As we have
mentioned earlier, the shifted dilaton remains invariant under duality and
$O(d,d)$ transformations. When we consider $\Omega$ transformation the shifted
dialaton is
$\tilde \phi = 2\Phi + 2\sigma - {{1\over 2}} ln g_{\theta \theta} $; whereas
the shifted dilaton ( which is the same as the original one ) is given by

$$ \tilde \phi = 2\Phi + \sigma - {{1\over 2}} ln detG' \eqno (22) $$

\noindent where the $\sigma$ comes from $g_{44}$ which remains unaffected under
the $\Omega$ transformarion. Similarly, we can see that under the $\tilde
\Omega$ transformation, the shifted dilaton is to be computed as we did in eq.
(22).

 One could perform other $O(3,3)$ transformations, for example on the $3-5$
plane to generate other background configurations. It is evident that, if we
 perform two $O(3,3)$ transformations say $\Omega_1$ followed by another,
$\Omega_2$ then the transformed $M$-matrix will be $\bar M$ ={ \Omega_2}^T
{\Omega_1}^T M \Omega_1 \Omega_2$. In fact, similar transformations were
employed in reference 4  to show how the Nappi-Witten cosmological model be
made singularity free by implementing such rotations.

We conclude this note with following remarks: we have shown how new backgrounds
can be generated through the noncompact symmetry transformations.
Since, $x_3$ is the internal compactified dimension, the $B_{\theta 3}$
component has the interpretation of being a gauge field such that
the vector potential has only nonzero $\theta$ component. Moreover, we observe
that the potential depends only on $r$. Therefore, the only surviving field
strength is $F_{r\theta} = \partial_r A_{\theta} - \partial_{\theta} A_r$. It
is evident from the form of the potential (see eq.17) that the vector potential
goes to a
constant value for large $r$ and the field strength falls off as $1 \over
{r^3}$ for large $r$. Since the vector potential arises from a component of the
antisymmetric tensor field as consequence of compactification, it is like an
axial vector. If we recall transformation of the vector potentials, arising out
of
compactifications, we notice that $\cal A \rightarrow \eta \cal A $ under
duality as is evident from equation (7). Therefore, with a sequence of $O(d,d)$
transformations we can generate both vector and axial vector type of Abelian
gauge fields.

It will be interesting to investigate how one generates new set of backgrounds
configurations for other solitonic solutions of the string effective action.

{\bf Acknowledgment:} I am grateful to Professor M. J. Duff for very
illuminating discussions on the solitonic solutions, explaining string/string
duality to me and carefully reading the manuscript. It is a pleasure to
acknowledge gracious hospitality of ProfessorT. W. B. Kibble and the Isaac
Newton Institute for Mathematical Sciences during the activities on Topology
and Defects.

\vfil
\eject

\def \np {{\it Nucl. Phys. }}
\def \pl {{\it Phys. Lett. }}
\def \prl {{\it Phys. Rev. Lett. }}
\def \pr {{\it Phys. Rev. }}

\noindent {\bf References:}

\item {[1]} K. Kikkawa and M. Yamasaki, {\it Phys. Lett.} {\bf 149B}
(1984) 357; N. Sakai and I. Senda, {\it Prog. Theor. Phys.} {\bf 75}
(1986) 692; T. Busher, {\it Phys. Lett.} {\bf 194B} (1987) 59;
{\it Phys. Lett.} {\bf 201B} (1988) 466;
{\it Phys. Lett.} {\bf 159B} (1985) 127; V. Nair, A. Shapere,
A. Strominger, and F. Wilczek, {\it Nucl. Phys.} {\bf B287} (1987) 402;
A. Giveon, E. Rabinovici,
and G. Veneziano, {\it Nucl. Phys.} {\bf B322} (1989) 167;
M. J. Duff, \np {\bf B335} (1990) 610,

 A. A. Tseylin and C. Vafa, {\it Nucl. Phys.} {\bf B372}
(1992) 443;
A. A. Tseytlin, {\it Class. Quan. Gravity} {\bf 9} (1992) 979. For recent
reviews see A. Sen Int. J. Mod. Phys. A (to appear) and A. Giveon, M. Porrati
and E. Ravinovici, Phy. Reports C ( to appear).

\item {[2]} K. A. Meissner and G. Veneziano, {\it Phys. Lett.} {\bf 267B}
(1991) 33; G. Veneziano, {\it Phys. Lett.} {\bf 265B} (1991) 287,
 K. A. Meissner and G. Veneziano, {\it Mod. Phys. Lett.} {\bf A6}
(1991) 3397.
\item {[3]} M. Gasperini, J. Maharana and G. Veneziano, {\it Phys. Lett.} {\bf
272B} (1991) 167, M. Gasperini and G. Veneziano, {\it Phys. Lett.} {\bf 277B}
(1991) 256.
\item {[4]} M. Gasperini. J. Maharana and G. Veneziano, {\it Phys. Lett.} {\bf
296} (1992) 51.
\item {[5]} A. Sen, \pl {\bf 271B} (1991) 295;
A. Sen, \pl {\bf 274B} (1991) 34;
S. F. Hassan and A. Sen, \np {\bf B375} (1992) 103.

\item {[6]} J. Scherk and J. H. Schwarz, \np {\bf B153} (1979) 61.
\item {[7]} J. Maharana and J. H. Schwarz, \np {\bf B390} (1993) 3. For earlier
work see S. Ferrara, C. Kounnas and M. Porrati, \pl {\bf B181} (1986) 263.
\item {[8]} M. J. Duff, Clas. and Quantum Grav. {\bf 5} (1988)
\item {[9]} A. Strominger, \np {\bf 343} (1990) 167.
\item {[10]} M. J. Duff and J. X. Lu, \np {\bf B354} (1991) 129.
\item {[11]} M. J. Duff and J. X. Lu, \np {\bf B354]} (1991) 141.
\item {[12]} C. G. Callan, J. A. Harvey and A. Strominger, \np {\bf B359}
(1991) 611; \np {\bf B367} (1991) 60.
\item {[13]} M. J. Duff and J. X. Lu, \prl {\bf 66} (1991) 1402.
\item {[14]} C. Montonen and D. Olive, \pl {\bf 72B} (1977) 117; P. Goddard, J.
Juyts and D. Olive \np {\bf B125} (1977) 1;
\item {[15]}  H. Osborn, \pl {\bf 83B} (1979) 321

\item {[16]} M. J. Duff, P. Howe, T. Inami and K. S. Stelle \pl {\bf 191B}
(1987) 70; M. J. Duff, T. Inami, C. N. Pope, E. Sezgin and K. S. Stelle \np
{\bf 297} (1988); K. Fujikawa and J. Kubo, \np {bf B365} (1991) 208.
\item {[17]} M. J. Duff and R. R. Khuri, \np {\bf B411} (1994) 473; M. J. Duff,
R. R. Khuri, R. Minasian and J. Rahmfeld, \np {\bf B418} (1994) 195.
\item {[18]} B. R. Greene, A. Shapere, C, Vafa and S. T. Yau, \np {\bf B337}
(1990) 1.
\item {[19]} A. Dhabolkar, G. Gibbons, J. A. Harvey and F. Ruiz Ruiz, \np {\bf
B340} (1990).
\item {[20]} S. P. Khastgir and J. Maharana, \pl {\bf 301B} (1993) 191; \np
{\bfB406} (1993) 162.
\item {[21]} A. Font, L. Ibanez and F. Quevedo, \pl {\bf 249B} (1990) 35; S.-J.
Rey, \pr {\bf D43} (1991) 526; A. Sen \np {\bf B404} (1993) 109; \pl {\bf 303B}
(1993) 22; {\ Mod.Phys. Lett.} {\bf A8} (1993) 2023; J. H. Schwarz and A. Sen
\pl {\bf 329B} (1994) 105, A. Sen, \pl {\bf 329B} (1994) 217.
\item {[22]} J. H. Schwarz and A. Sen \np {\bf B411} (1994) 35.
\item {[23]} P. Binetruy, preprint NSF-ITP-93-60.
\item {[24]} C. Vafa and E. Witten preprint HUTP-94-A017, N. Seiberg and E.
Witten, preprint RU 94-52, RU-94-60; L. Girardello, A. Giveon, M. Porrati and
A. Zafaroni, preprint NYU-TH 94/05/02, J. Horne and G. Moore, preprint
YPT-P2-94.
\item {[25]} A. Sen preprint TIFR-TH-94-19; G. Segal ( to appear) and E. Witten
(unpublished).

\end